\documentclass{article}
\usepackage{spconf,amsmath,graphicx, multirow, url}

\usepackage{enumitem}
 \ninept
\setlist{nosep, leftmargin=14pt}

\title{Lightweight and Interpretable Left Ventricular Ejection Fraction Estimation using Mobile U-Net}
\name{Meghan Muldoon, Naimul Khan}
\address{Toronto Metropolitan University, Toronto, Canada}
\begin{document}

%\ninept
%
\maketitle
\begin{abstract}
Accurate LVEF measurement is important in clinical practice as it identifies patients who may be in need of life-prolonging treatments. This paper presents a deep learning based framework to automatically estimate left ventricular ejection fraction from an entire 4-chamber apical echocardiogram video. The aim of the proposed framework is to provide an interpretable and computationally effective ejection fraction prediction pipeline. A lightweight Mobile U-Net based network is developed to segment the left ventricle in each frame of an echocardiogram video. An unsupervised LVEF estimation algorithm is implemented based on Simpson's mono-plane method. Experimental results on a large public dataset show that our proposed approach achieves comparable accuracy to the state-of-the-art while being significantly more space and time efficient (with 5 times fewer parameters and 10 times fewer FLOPS).

\end{abstract}
\section{Introduction}
\label{sec:intro}

Left ventricular ejection fraction (LVEF) is one of the most important metrics in assessing cardiac function. LVEF represents the difference in left-ventricular end-systolic volume and left-ventricular end-diastolic volume \cite{LVEF}. Accurate calculation of LVEF is important in clinical practice as it identifies patients who may be in need of life-prolonging treatments. It can also be used as a prognostic factor to predict adverse outcomes in individuals due to cardiovascular disease \cite{LVEF}.

Clinicians can estimate LVEF manually by analyzing 2D echocardiogram videos across multiple cardiac cycles. Echocardiogram studies offer several advantages compared to other medical imaging modalities as they are low cost and non-invasive. However, echocardiogram imaging also suffers several limitations including complex analyses during evaluation, high operator subjectivity, wide operation ranges and observer variability \cite{Zhou:2021}.

Automated LV segmentation in echocardiogram studies is a challenge due to heart structure motion during the cardiac cycle, respiratory interference and noise/artifacts inherent to ultrasound imaging \cite{Zhou:2021}. Additionally, large labeled medical datasets are uncommon and most segmentation methods must account for limited data. Recently proposed frameworks include active appearance models, level sets and computer vision based frameworks \cite{Jafari:2019}. Many deep learning based left ventricular segmentation applications have also been developed.

Multi-domain regularized fully CNNs have shown success in segmenting anatomical structures in ultrasound images. \cite{Chen:2016}. U-Net architectures have also been applied to LV segmentation tasks, however it is not guaranteed that they will produce anatomically plausible masks across entire videos. \cite{Smistad:2017}. 

Simpson's method is the recommended procedure for manually assessing LVEF in a single plane echocardiogram videos \cite{LVEF}. The method requires accurate measurements of LV areas throughout a cardiac cycle, averaged over multiple heart beats. This calculation can be time-consuming for clinicians as it requires precise annotations on multiple cardiac frames, which are often noisy and contain artifacts. Many methods have been proposed in recent literature to automate this process.

Zhang et al. propose a model to interpret echocardiogram studies which implements a U-Net architecture for LV segmentation and estimates LVEF using Simpson's method. \cite{Zhang:2018}. Jafari et al. propose a similar method, in which the LV and landmarks of interest are segmented using a modified U-Net with adversarial training \cite{Jafari:2019}. Other proposed architectures use deep learning for both segmentation and LVEF estimation. Ouyang et al. propose a model which uses a DeepLabv3 framework to segment the LV and a 3D CNN for LVEF estimation\cite{Ouyang:2020}. A simplified version of this framework is proposed by Lagopoulos et al. in which a regression model is trained to estimate LVEF using geometric features of segmentation masks \cite{Lagopoulos:2022}. These frameworks perform on par with human experts, however they are computationally complex and often contain hundreds of millions of parameters. This means that they may not be able to run on point of care ultrasound devices, where computational resources are limited \cite{Jafari:2019}.

Attention based methods have also been proposed to estimate LVEF\cite{Renaud2021} \cite{Fazry2022} \cite{Muhtaseb2022}. Although these "black-box" methods perform well they are not considered interpretable because they have no intermediate steps that can be verified by a clinician \cite{TrustworthyAI}. Fully autonomous frameworks are currently not recommended in clinical practice because they lack transparency.

This paper presents a deep learning based framework to automatically estimate LVEF from an entire 4-chamber apical echocardiogram video. The proposed framework aims to be computationally efficient while providing visually interpretable intermediate results for clinical transparency.  We propose a lightweight Mobile U-Net based network with tracking to segment left ventricle in each frame of an echocardiogram video. An LVEF estimation algorithm is implemented based on clinical methods. Experimental results on a large public dataset show that our proposed approach achieves comparable accuracy to the state-of-the-art while being space and time efficent. 

\section{Methodology}
\label{sec:method}

\subsection{Left Ventricular Segmentation}
Accurate segmentation of the LV is important for estimating LVEF and assessing cardiac function. The presence of a segmentation component also adds increased interpretability to LVEF estimation frameworks. A frame level left ventricular segmentation network was created and trained on the ground truth end-diastolic (ED) and end-systolic (ES) tracings from the EchoNet dataset\cite{EchoNet}. 

\subsubsection{Mobile U-Net}

\begin{figure}[htb]

\begin{minipage}[b]{1.0\linewidth}
  \centering
  \centerline{\includegraphics[width=0.85\textwidth]{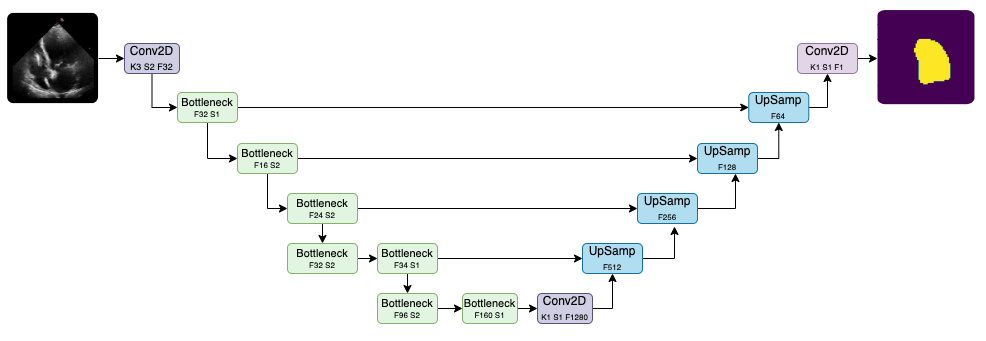}}

\end{minipage}
\caption{Full Mobile U-Net architecture for left ventricular segmentation. $F$:filters, $S$: stride, $K$: kernel size}
\label{fig:mu-net}
\end{figure}

Traditional U-Net architectures has approximately 30 million parameters and have shown exceptional performance on many medical image segmentation tasks. A U-Net model consists of two components; a contracting (encoder) pathway and an expanding (decoder) pathway \cite{U-Net}. The contracting path is made up of a generic convolutional network architecture which is responsible for producing feature maps at various resolutions. The expanding pathway up-samples the features and concatenates them with the corresponding feature map from the contracting path. To create a network with fewer parameters than a traditional U-Net, the contracting pathway in the proposed framework was replaced by a lightweight MobileNetV2 based architecture illustrated in Figure \ref{fig:mu-net}.

Mobile U-Net architectures have shown comparable performance to U-Net with fewer than half of the parameters \cite{MU-Net}\cite{RMU-Net}. The encoder pathway is composed of a convolutional layer followed by 7 Bottleneck Residual blocks, as described in \cite{MobileNet}. The number of filters, kernel size, expansion factor and stride for each layer are kept the same as described in \cite{MobileNet}. The decoder portion of the  U-Net architecture was maintained. The final Mobile U-Net segmentation framework is composed of approximately 12 million parameters.

\subsubsection{Mask Track}
\begin{figure}[b]
\begin{minipage}[b]{1.0\linewidth}
  \centering
  \centerline{\includegraphics[width=0.65\textwidth]{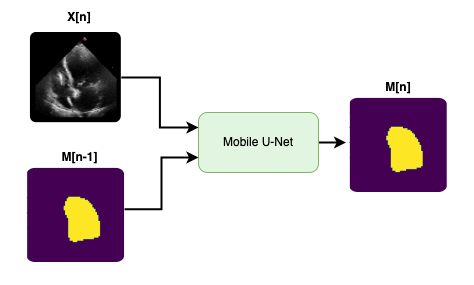}}
\end{minipage}
\caption{Given an input frame $X[n]$ and a mask estimate from the previous frame $M[n-1]$, MaskTrack predicts a refined mask for the current frame $M[n]$}
\label{fig:masktrack}
\end{figure}
Though we have access to a large dataset of echocardiogram videos, annotated masks are available for less than 1\% of the training frames. Instead of focusing on LV segmentation as a per-frame problem, we attempt to treat the problem as a video object segmentation task in which a single mask is tracked across an entire echocardiogram video. We propose modifying Mobile U-Net to behave as a MaskTrack architecture based on \cite{MASKTRACK}. MaskTrack is a guided instance segmentation framework that can be appended to any convolutional network. The model learns to leverage the previous mask with the spatial information in the input frame to produce smooth video segmentation, shown in Figure \ref{fig:masktrack}.

Any existing convolutional network which segments three channel (RGB) images can be converted to a MaskTrack framework by expanding it to take RGB+mask input. The only change made to Mobile U-Net is an added channel in the first convolutional layer of the network, which takes the mask channel as input.  The model is trained on frames with annotated ground truths. A simulated previous frame is generated for each input image by applying random affine transformations and non-rigid deformations to the annotation. These transformations aim to simulate the movement of the left-ventricle across frames. 
\begin{figure}[h]

  \centerline{\includegraphics[width=0.40\textwidth]{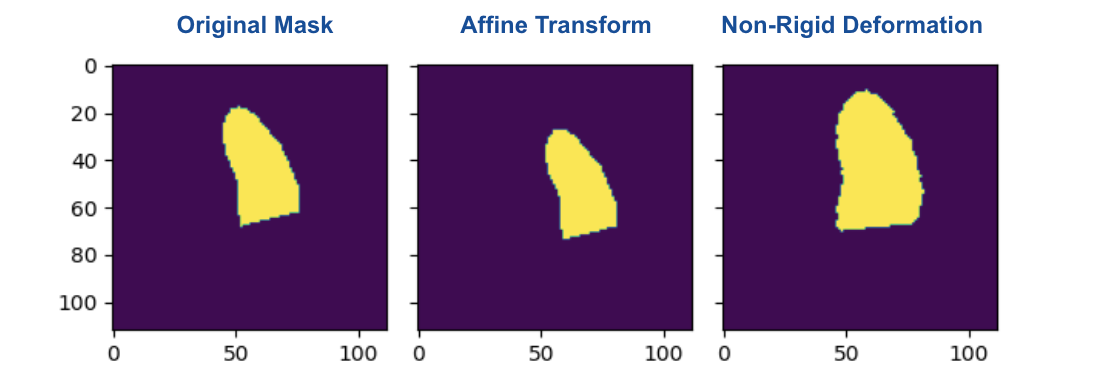}}

\caption{Data augmentation for an LV ground truth mask}
\label{fig:augs}
\end{figure}
Affine transforms consist of random scaling $\pm$10\% of the original mask size and translation of $\pm$10\% of the frame size.  Non-rigid deformations are performed via thin-plate spline (TPS) mapping. The thin-plate spline warp is approximated by sampling a subset of 5 points on the contour and randomly shifting them $\pm$ 10\% of the mask width and height. The TPS coefficients are calculated and then applied to the remaining points on contour \cite{tps-approx}. Figure \ref{fig:augs} shows the warps applied to a random mask. Thin plate spline transformations are well suited to LV data augmentation as they have been used to model biomedical images and have been used successfully in data augmentation frameworks for MRI images \cite{tps} \cite{Lin2019}.
\subsection{Unsupervised Estimation of LVEF}
The LVEF framework estimates ejection fraction based on the predicted masks from the segmentation network. LV volume is calculated for each frame using Simpson's monoplane (area length) method. Beat analysis is used to identify individual cardiac cycles from LV volumes and estimate the ejection fraction for each cardiac cycle. The final LVEF estimation is calculated as the average of EF across multiple heartbeats. 

\subsubsection{Volume Estimation}
The area length method (Equation \ref{eq1}) is used to estimate left ventriclar volume based on the geometry of the segmentation mask. The length and area of the mask is calculated for each segmentated left ventricle. The volume estimation framework implements the algorithm described by Moradi et al. which imitates the process that clinicians use to find the long axis length in 4-chamber echocardiograms \cite{Moradi2019} \cite{LVEF}. For each frame the procedure is as follows:
\begin{enumerate}
	\item Identify contour of mask from the segmentation network.
	\item Find the minimum enclosing triangle containing the contour using O'Rourkes algorithm \cite{mintri}.
	\item For each triangle vertex find the closest point on the contour. The two closest points on the contour mark the mitral valve annulus. The remaining point marks the LV apex
	\item The midline is defined as the line between the apex point and the midpoint of the mitral valve annulus.
	\item The midline is intersected with the contour.
	\item LV length is defined as the distance between the apex and the intersection point.
	\item The LV volume estimated using the ellipsoid model equation \cite{LVEF}.  Where $A$ and $L$ represent the mask area and LV length respectively. These values are both measured in pixels.
	\begin{equation}
		\label{eq1}
		Volume = \frac{8 A^2}{3 \pi L}
	\end{equation}
\end{enumerate}

\begin{figure}[htb]
\begin{minipage}[b]{1.0\linewidth}
  \centering
  \centerline{\includegraphics[width=0.35\textwidth]{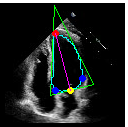}}
\end{minipage}
\caption{Cyan: mask contour. Green: minimum enclosing triangle. Blue: bounds of the mitral valve annulus. Red: LV apex. Yellow: intersection of midline and contour. Magenta: LV length measurement}
\label{fig:lv_length}
\end{figure}

Figure \ref{fig:lv_length} illustrates this procedure on a sample frame in the EchoNet Dynamic Dataset.

\subsubsection{Beat Analysis}
\label{beat-analysis}
\begin{figure}[h]

  \centering
  \centerline{\includegraphics[width=0.55\textwidth]{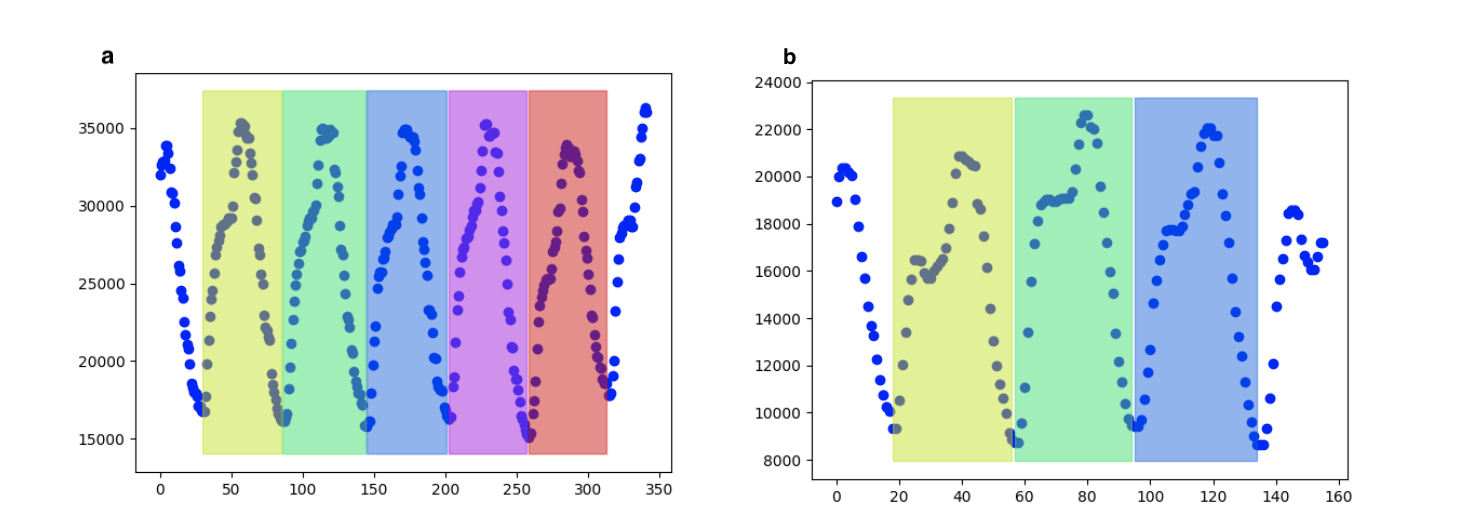}}
\caption{a) Volume signal with five identified cardiac cycles. b) Volume signal with three identified cardiac cycles}
\label{fig:beat-anal}

\end{figure}

Ejection fraction is predicted for the entire video by averaging multiple beat-level EF estimations. This allows the framework to  accept videos of varying length containing any number of cardiac cycles. The volumes for an input clip are passed through a median filter to remove outliers. Peaks and troughs are identified in the filtered volume signal, demonstrated in Figure \ref{fig:beat-anal}. Peaks indicate maximum LV volume (end-diastole) and indicate minimum LV volume (end-systole). Each cardiac cycle is defined between adjacent troughs and the end-diastolic volume and end-systolic volumes are determined. Ejection fraction for each cycle ($k$) is calculated using Equation 2.

\begin{equation}
	EF_k = \frac{V_{ED} - V_{ES}}{V_{ED}}  
\end{equation}
\begin{equation}
	EF = \frac{1}{N} \sum_{k=1}^{N} EF_k
\end{equation}

The final estimate of ejection fraction for the video is the mean of the ejection fraction in each beat, Equation 3. This process is similar to the beat-to-beat assessment method described in Ouyang et. al \cite{Ouyang:2020}, however this method calculates each EF directly based on the segmentation volumes rather than averaging EF volumes estimated by an external network. This concept aligns with clinical guidelines which recommend averaging ejection fraction across multiple beats \cite{Lang2015}.

\section{Experiments and Results}
\label{sec:exp}

\subsection{Dataset}
\label{dataset}
The EchoNet Dynamic dataset was used to train and evaluate the proposed framework \cite{EchoNet}. It consists of 10 030 apical-4 chamber echocardiogram videos of healthy hearts and various pathologies. The videos vary in length, frame rate and image quality. Each video has a set of associated measurements obtained by a clinician, including end-diastolic volume, end- systolic volume and LVEF. Tracings of the left ventricle are provided for the ED and ES frames of one cardiac cycle in each video. The dataset is divided into training, validation and testing subsets, with 75\% of videos in the training set and 12.5\% of data in the validation and test set respectively. The distribution of data is similar in each cohort and dataset summary statistics are reported in \cite{EchoNet}.

\subsection{Results}
Several experiments were conducted to assess the effectiveness of each component in the proposed framework.  MU-Net outperformed MUNet + MaskTrack in all evaluation methods. 

\begin{table}
		\caption{Metrics comparing MU-Net and MU-Net + MaskTrack}
		\label{table-ef}
		\centering
		%\resizebox{\columnwidth}{!}{
				\begin{tabular}{|c|c c|}

						\hline
						Framework&  LVEF MAE  & DSC \\ 
						\hline\hline

						MU-Net  & \textbf{ 6.61 }& \textbf{0.905} \\
						MU-Net + MaskTrack  &  8.24& 0.850\\
						
						\hline
						
					\end{tabular}%
			
	\end{table}

\subsubsection{Segmentation}
Dice coefficent (DSC) was used as a metric to assess the segmentation frameworks. Dice coefficent is a pixel-wise measure of agreement between a predicted mask ($X$) and ground-truth ($Y$). The formula is given by:

\begin{equation}
	DSC = \frac{2|X \cap Y|}{|X| + |Y|} 
\end{equation}

\begin{figure}[h]

  \centering
  \centerline{\includegraphics[width=0.48\textwidth]{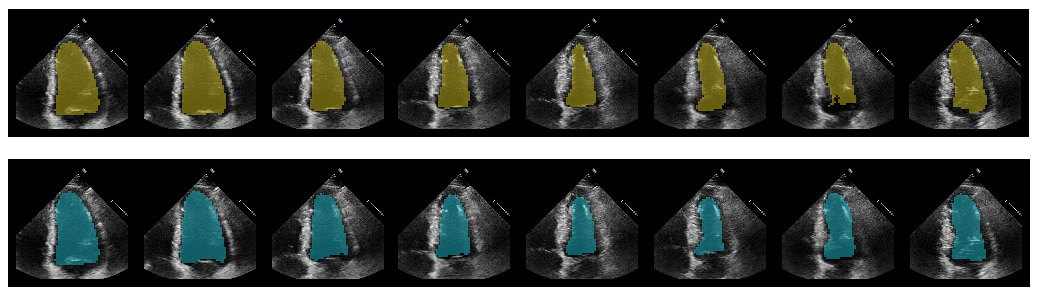}}
\caption{Top: LV segmentations using MU-Net. Bottom: LV segmentations using Mobile U-Net + Mask track }
\label{fig:videos}

\end{figure}

\begin{figure}[h]
  \centering
  \centerline{\includegraphics[width=0.35\textwidth]{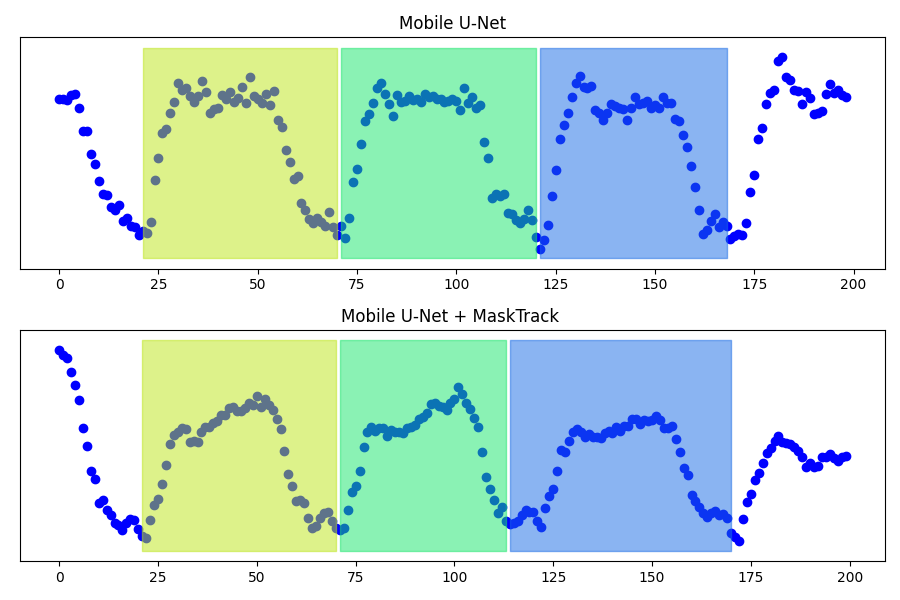}}
\caption{Volumes signals for Mobile U-Net and Mobile U-Net + MaskTrack}
\label{fig:comparevols}
\end{figure}

A ground truth tracing is provided on one systolic and diastolic frame for each video in the EchoNet Dynamic dataset. This results in a test set of 2552 frames to assess the segmentation performance of the network. Table \ref{table-ef} shows the DSC for MU-Net and MU-Net + MaskTrack. MU-Net + Mask Track has a lower average DSC on the ground truth frames than MU-Net, likely because error accumulates over time.  Well performing examples of MaskTrack produce visually smoother video segmentation than MU-Net (Figure \ref{fig:videos}), however the performance on intermediate frames cannot be assessed because ground truth frames are only provided for end diastole and end systole. Figure \ref{fig:comparevols} illustrates that MU-Net + Mask Track produces a smoother volumes graph than MU-Net.  

\subsubsection{Ejection Fraction Estimation}
The performance of the framework was assessed by taking the mean-absolute-error between the predicted ejection fraction and the ground truth for each video. Mean absolute error is taken as:
\begin{equation}
	MAE = \frac{1}{N} \sum_{k=1}^N|\text{EF predicted}_k - \text{EF true}_k|
\end{equation}

The best performance was achieved by MU-Net when estimating LVEF as the mean of the calculated ejection fraction for each identified cardiac cycle. 

\subsubsection{LVEF Classification}
In practice a clinician will often classify LVEF measurements into the ranges: reduced LVEF (rEF; LVEF $<$ 0.40),  moderately reduced LVEF (mrEF; LVEF 0.40-0.49) and  preserved LVEF (pEF; LVEF  $>$ 0.50). Clinicians are required to make these classifications when diagnosing and treating heart failure \cite{Lang2015}. To gauge the performance of the model in assigning an echocardiogram video to an LVEF range, the proposed framework was applied to an LVEF classification task. A label was assigned to each video in the test set based on the predicted ejection fraction. 

\begin{table}
	\renewcommand{\arraystretch}{1.3}
	\caption{Classification performance of the proposed frameworks. $F1_{\mu}$: F1 score obtained by micro-averaging based on class frequency.  $F1_{m}$: F1 score obtained by taking the unweighted mean of all per-class F1 scores}
	\label{table-cla}
	\centering
	\resizebox{\columnwidth}{!}{
		\begin{tabular}{|c| c c c c|}

			\hline
			Framework& $F1_\mu $ & $F1_m$  & $\text{Recall}_m$& $\text{Precision}_m$\\ 
			\hline\hline
			MU-Net  & \textbf{0.828}& \textbf{0.621} & \textbf{0.593} & \textbf{0.671}\\
			MU-Net + MaskTrack  &0.777 & 0.538& 0.517&0.578 \\
			
			\hline
			
		\end{tabular}%
	}
\end{table}

Table \ref{table-cla} displays the results of LVEF classification using MU-Net and MU-Net+MaskTrack. The Mobile U-Net framework without MaskTrack performed best. This framework was successful at predicting the rEF and pEF classes, with F1 scores of 0.92 and 0.65 respectively. The mrEF class had a far worse F1 score of 0.29. From inspection of the confusion matrix (Table \ref{table-cf})  we can see that mrEF cases are often mislabeled as pEF by the proposed framework. 

	\begin{table}[!ht]
	
		\centering
		\caption{Confusion matrix for the LVEF classification using Mobile U-Net}
		\label{table-cf}
		\renewcommand{\arraystretch}{1.2}

		\begin{tabular}{ll|l|l|l|l|l|}
			
			\multicolumn{2}{c}{}&   \multicolumn{3}{c}{\textbf{Predicted Class}}\\
			\multicolumn{2}{c}{}&\multicolumn{3}{c}{{\rotatebox[origin=c]{0}{pEF}
				} {\rotatebox[origin=c]{0}{rEF}
				} {\rotatebox[origin=c]{0}{mrEF}
			}}\\
			\cline{3-5}
			\multirow{3}{*}{{\rotatebox[origin=c]{90}{\textbf{True Class}}
			}} & 
			pEF&932 & 6 &53 \\ \cline{3-5}
			&   rEF&28& 86 &46\\ \cline{3-5}
			&   mrEF&76 & 11&38  \\ \cline{3-5}

		\end{tabular}
	\end{table}

\subsection{Comparison to State of the Art}

The proposed method was compared with six other LVEF estimation frameworks. Each framework in Table \ref{table-soa} recorded test metrics on the Echo-Net Dynamic dataset. Though the best performance of the proposed model is lower than the other frameworks, the reported MAE (6.61\%) is within interobserver variation (which ranges from 7.9\% to 13.9\%).

The proposed framework is more space and time efficient than existent architectures. Additionally, many proposed solutions implement a "black-box" approach, where LVEF is estimated directly from an echo clip with no intermediate steps \cite{Renaud2021}\cite{Fazry2022}\cite{Muhtaseb2022}. Though these methods produce accurate results, they may be difficult for clinicians to interpret. The proposed framework outputs the segmented mask video, allowing a clinician to verify the EF estimation by examining the accuracy of the masks.  

\begin{table}
	\renewcommand{\arraystretch}{1.2}
	\caption{Comparison of the proposed framework to other LVEF estimation frameworks.}
	\label{table-soa}
	\centering
		\resizebox{\columnwidth}{!}{
\begin{tabular}{|c| c c c c|} 
	
	\hline
	Model & MAE & RMSE  & Params&FLOPS\\ 
	\hline\hline
	Ouyang et al.\cite{Ouyang:2020} & 4.05 &	5.32 & 71.1M & 	91.97G\\
	Lagopolous et al.\cite{Lagopoulos:2022} & 5.08&6.92 & ~39.9M & - \\
	Reynaud et al.\cite{Renaud2021} & 5.95& 8.38&- &130.0G	\\
	Fazry et al.\cite{Fazry2022} & 5.72& 7.63& 49.7M& - \\
	Muhtaseb et al.\cite{Muhtaseb2022}&\textbf{3.95} & \textbf{5.17} &-& 19.61G \\
	Blaivas et al.\cite{Blaivas2022}&8.08 & 11.98 &138.4M& - \\
	MU-Net & 6.61  &8.91& \textbf{12.38M}& \textbf{2.06G} \\

	\hline
\end{tabular}
}
\end{table}

\section{Conclusion}
\label{sec:conc}
In this work we propose a lightweight framework for estimating LVEF in echocardiogram videos. The proposed framework provides visually interpretable intermediate results, compared to existing ”black-box” models. Though the proposed framework has higher MAE than some existent architectures it contains approximately 12 million parameters, while have hundreds of millions of parameters. Our lightweight model has the potential to be used in point of care ultrasound settings, where computational resources may be limited. 

Though the MaskTrack module did not improve the results of Mobile U-Net, it produced visually smooth video segmentation. In the future we would like to experiment with both the MU-Net and Masktrack on other biomedical video datasets. The code from this work is available at: \url{https://github.com/megmuldoon/lvef_estimation}.

% Below is an example of how to insert images. Delete the ``\vspace'' line,
% uncomment the preceding line ``\centerline...'' and replace ``imageX.ps''
% with a suitable PostScript file name.
% -------------------------------------------------------------------------

\vfill
\pagebreak

\section{Compliance with ethical standards}
\label{sec:ethics}

This study was conducted retrospectively using human subject data made available in open access by \cite{EchoNet}.

\section{Acknowledgments}
No funding was received for conducting this study. The authors have no relevant financial or non-financial interests to disclose

\bibliographystyle{IEEEbib}
\bibliography{strings,refs}

\end{document}